\documentclass[fleqn,usenatbib]{mnras}

\usepackage{newtxtext,newtxmath}

\usepackage[T1]{fontenc}

\DeclareRobustCommand{\VAN}[3]{#2}
\let\VANthebibliography\thebibliography
\def\thebibliography{\DeclareRobustCommand{\VAN}[3]{##3}\VANthebibliography}

\usepackage{graphicx}	
\usepackage{amsmath}	
\usepackage{rotating}

\usepackage{orcidlink}
\usepackage{xspace}
\usepackage{ulem}
\makeatletter 
  \patchcmd{\NAT@citex}
    {\@citea\NAT@hyper@{%
      \NAT@nmfmt{\NAT@nm}%
      \hyper@natlinkbreak{\NAT@aysep\NAT@spacechar}{\@citeb\@extra@b@citeb}%
      \NAT@date}}
    {\@citea\NAT@nmfmt{\NAT@nm}%
    \NAT@aysep\NAT@spacechar\NAT@hyper@{\NAT@date}}{}{}

  \patchcmd{\NAT@citex}
    {\@citea\NAT@hyper@{%
      \NAT@nmfmt{\NAT@nm}%
      \hyper@natlinkbreak{\NAT@spacechar\NAT@@open\if*#1*\else#1\NAT@spacechar\fi}%
        {\@citeb\@extra@b@citeb}%
      \NAT@date}}
    {\@citea\NAT@nmfmt{\NAT@nm}%
    \NAT@spacechar\NAT@@open\if*#1*\else#1\NAT@spacechar\fi\NAT@hyper@{\NAT@date}}
    {}{}
\makeatother

\newcommand{\thesan}{\textsc{thesan}\xspace}
\newcommand{\thzoom}{\mbox{\textsc{thesan-zoom}}\xspace}

\title[Overmassive black holes in the early Universe]{Overmassive black holes in the early Universe can be explained by gas-rich, dark matter-dominated galaxies}

\author[W. McClymont et al.]{%
William McClymont $\orcidlink{0009-0009-5565-3790}$,$^{1,2}$\thanks{E-mail: \href{mailto:wjm50@cam.ac.uk}{wjm50@cam.ac.uk} (WM)}
Sandro Tacchella $\orcidlink{0000-0002-8224-4505}$,$^{1,2}$
Xihan Ji $\orcidlink{0000-0002-1660-9502}$,$^{1,2}$
Rahul Kannan $\orcidlink{0000-0001-6092-2187}$,$^{3}$
\newauthor
Roberto Maiolino $\orcidlink{0000-0002-4985-3819}$,$^{1,2}$
Charlotte Simmonds $\orcidlink{0000-0003-4770-7516}$$^{1,2}$
Aaron Smith $\orcidlink{0000-0002-2838-9033}$,$^{4}$
Ewald Puchwein $\orcidlink{0000-0001-8778-7587}$,$^{5}$
\newauthor
Enrico Garaldi $\orcidlink{0000-0002-6021-7020}$,$^{6,7}$
Mark Vogelsberger \orcidlink{0000-0001-8593-7692},$^{8}$
Francesco D'Eugenio $\orcidlink{0000-0003-2388-8172}$,$^{1,2}$
Laura Keating $ \orcidlink{0000-0001-5211-1958}$,$^{9}$
\newauthor
Xuejian Shen $\orcidlink{0000-0002-6196-823X}$,$^{8}$
Bartolomeo Trefoloni $\orcidlink{0009-0007-3295-8669}$,$^{10,11}$
and Oliver Zier $\orcidlink{0000-0003-1811-8915}$$^{12,8}$
\\
\\
$^{1}$Kavli Institute for Cosmology, University of Cambridge, Madingley Road, Cambridge CB3 0HA, UK\\
$^{2}$Cavendish Laboratory, University of Cambridge, 19 JJ Thomson Avenue, Cambridge CB3 0HE, UK\\
$^3$ Department of Physics and Astronomy, York University, 4700 Keele Street, Toronto, ON M3J 1P3, Canada \\
$^4$ Department of Physics, The University of Texas at Dallas, Richardson, TX 75080, USA \\
$^5$ Leibniz-Institut f\"ur Astrophysik Potsdam, An der Sternwarte 16, 14482 Potsdam, Germany \\
$^6$ Kavli IPMU (WPI), UTIAS, The University of Tokyo, Kashiwa, Chiba 277-8583, Japan \\
$^7$ Center for Data-Driven Discovery, Kavli IPMU (WPI), UTIAS, The University of Tokyo, Kashiwa, Chiba 277-8583, Japan \\
$^{8}$ Department of Physics, Kavli Institute for Astrophysics and Space Research, Massachusetts Institute of Technology, Cambridge, MA 02139, USA \\
$^9$ Institute for Astronomy, University of Edinburgh, Blackford Hill, Edinburgh, EH9 3HJ, UK \\
$^{10}$ Dipartimento di Fisica e Astronomia, Università di Firenze, via G. Sansone 1, 50019 Sesto Fiorentino, Firenze, Italy \\
$^{11}$ INAF – Osservatorio Astrofisico di Arcetri, Largo Enrico Fermi 5, I-50125 Firenze, Italy \\
$^{12}$ Center for Astrophysics $|$ Harvard $\&$ Smithsonian, 60 Garden Street, Cambridge, MA 02138, USA
}

\date{Accepted XXX. Received YYY; in original form ZZZ}

\pubyear{\the\year{}}

\begin{document}
\label{firstpage}
\pagerange{\pageref{firstpage}--\pageref{lastpage}}
\maketitle

\begin{abstract}
\textit{JWST} has revealed the apparent evolution of the black hole (BH)--stellar mass ($M_\mathrm{BH}$–$M_\mathrm{\ast}$) relation in the early Universe, while remaining consistent with the BH--dynamical mass ($M_\mathrm{BH}$–$M_\mathrm{dyn}$) relation. We predict BH masses for $z>3$ galaxies in the high-resolution \thzoom simulations by assuming the $M_\mathrm{BH}$–$M_\mathrm{dyn}$ relation is fundamental. Even without live BH modelling, our approach reproduces the \textit{JWST}-observed $M_\mathrm{BH}$ distribution, including overmassive BHs relative to the local $M_\mathrm{BH}$–$M_\mathrm{\ast}$ relation. We find that $M_\mathrm{BH}/M_\mathrm{\ast}$ declines with $M_\mathrm{\ast}$, evolving from $\sim$0.1 at $M_\mathrm{\ast}=10^6\,\mathrm{M_\odot}$ to $\sim$0.01 at $M_\mathrm{\ast}=10^{10.5}\,\mathrm{M_\odot}$. This trend reflects the dark matter ($f_\mathrm{DM}$) and gas fractions ($f_\mathrm{gas}$), which decrease with $M_\mathrm{\ast}$ but show little redshift evolution down to $z=3$, resulting in small $M_\mathrm{\ast}/M_\mathrm{dyn}$ ratios and thus overmassive BHs in low-mass galaxies. We use \texttt{Prospector}-derived stellar masses and star-formation rates to infer $f_\mathrm{gas}$ across 48,022 galaxies in JADES at $3<z<9$, finding excellent agreement with our simulation. Our results demonstrate that overmassive BHs would naturally result from a fundamental $M_\mathrm{BH}$–$M_\mathrm{dyn}$ relation and be typical of the gas-rich, dark matter-dominated nature of low-mass, high-redshift galaxies. Such overmassive BHs may strongly influence early galaxy formation, and we caution that our approach does not include the self-consistent BH-galaxy co-evolution required for a complete understanding.
\end{abstract}




\section{Introduction}
\label{sec:Introduction}

Supermassive black holes (BHs) appear nearly ubiquitous within the nuclei of local galaxies \citep{Heckman:2014aa,Greene:2020aa}. Their properties are tightly correlated with the properties of their host galaxies, which, alongside direct evidence for dramatic feedback driven by active galactic nuclei \citep[AGN;][]{Cicone:2012aa,Cicone:2014aa,Maiolino:2012aa,Genzel:2014aa}, has been taken as evidence for a co-evolution of BHs and their host galaxies \citep[e.g.,][but see also, \citealt{Shields:2003aa, Jahnke:2009aa, Merloni:2010aa, Bennert:2011aa, Shen:2015aa, Ding:2020aa, Suh:2020aa, Farrah:2023aa, Li:2023aa}]{Kormendy:2013aa, Greene:2020aa}. An abundance of models have explored this co-evolution through the lens of galaxy and BH mergers and feedback processes which jointly regulate star formation and BH accretion \citep{Sijacki:2009aa,Volonteri:2010aa,Valiante:2016aa,Inayoshi:2020aa,Koudmani:2022aa,Zhu:2022aa,Bennett:2024aa,Farcy:2025aa,Sanati:2025aa}. However, the origins and early evolution of BHs remains uncertain, particularly in the context of the extreme BH masses ($M_\mathrm{BH}\gtrsim10^9\,\mathrm{M_\odot}$) required to explain the luminous quasars already in place at $z\approx6$ \citep{Banados:2018aa,Wang:2020aa,Fan:2023aa}. Various mechanisms have been proposed to facilitate the growth of such massive BHs at early cosmic times, including heavy seeding scenarios, the rapid merging of light seeds, super-Eddington accretion, and ``dark seeding'' \citep{Ferrara:2014aa,Banik:2019aa,Inayoshi:2020aa,Xiao:2021aa,Sassano:2023aa,Narayan:2023aa,Schneider:2023aa,Jiang:2025aa,Shen:2025ab}.

The study of the high-redshift BHs has recently been revolutionized by the \textit{James Webb Space Telescope (JWST)}, which has revealed an abundance of AGN in the early Universe \citep{Kocevski:2023aa,Kocevski:2025aa,Ubler:2023aa,Ubler:2024aa,Ubler:2024ab,Harikane:2023aa,Perna:2023aa,Perna:2025aa,Larson:2023aa,Maiolino:2024aa,Maiolino:2024ac,Maiolino:2024ab,Matthee:2024aa,Greene:2024aa,Juodzbalis:2024aa,Juodzbalis:2024ab,Scholtz:2025aa}. These AGN have been identified through various approaches, including photometric selection \citep{Furtak:2023aa,Juodzbalis:2023aa,Onoue:2023aa,Yang:2023ab,Barro:2024aa} and the use of new emission line diagnostics appropriate for the low-metallicity, high-redshift regime \citep{Mazzolari:2024aa,Mazzolari:2025aa,Chisholm:2024aa,Scholtz:2025aa}. Of particular interest are those AGN that have been identified as Type 1 through their broad Balmer line emission \citep{Kocevski:2023aa,Ubler:2023aa,Harikane:2023aa,Maiolino:2024aa,Maiolino:2024ac,Tripodi:2024aa}. Through the use of local virial relations, it has been possible to calculate the single-epoch BH masses, $M_\mathrm{BH}$, for this subset of AGN \citep{Kocevski:2023aa,Ubler:2023aa,Harikane:2023aa,Maiolino:2024ac,Tripodi:2024aa,Juodzbalis:2025aa,Taylor:2025aa}, though the reliability of such measurements is debated \citep{Juodzbalis:2025aa,Naidu:2025aa,Rusakov:2025aa}. For many of these objects, careful spectral and imaging decomposition has been carried out to measure the stellar mass, $M_\mathrm{\ast}$, of the host galaxy \citep{Harikane:2023aa,Maiolino:2024ac,Juodzbalis:2025aa}.

The measurement of BH and host galaxy masses for faint sources at early cosmic epochs has facilitated the testing of BH-galaxy scaling relations at lower masses and higher redshifts, which has revealed an apparent evolution of the $M_\mathrm{BH}$–$M_\mathrm{\ast}$ relation. Nearly all host galaxies measured at $z>3$ with stellar masses below $10^9\,\mathrm{M_\odot}$ show BHs which are overmassive compared to local $M_\mathrm{BH}$–$M_\mathrm{\ast}$ relations by at least $1-2$\,dex \citep{Ubler:2023aa,Ubler:2024ab,Harikane:2023aa,Kokorev:2023aa,Carnall:2023aa,Pacucci:2023aa,Maiolino:2024aa,Maiolino:2024ac,Furtak:2024aa,Juodzbalis:2024ab,Natarajan:2024aa}. 
While selection bias certainly plays a role in this offset, as more massive black holes are easier to detect \citep{Juodzbalis:2024ab,Li:2025ab}, the existence of such overmassiveness could have key implications on the BHs early seeding and growth mechanisms \citep[e.g.][]{Schneider:2023aa,Silk:2024aa}. Interestingly, for a subset of the \textit{JWST}-discovered AGN where it has been possible to measure the dynamical mass of the host, $M_\mathrm{dyn}$, they are largely consistent with the locally derived $M_\mathrm{BH}$–$M_\mathrm{dyn}$ relation \citep{Maiolino:2024ac}. This finding indicates on one hand that the selection effect on the broad-lines flux cannot entirely explain the offset on the $M_\mathrm{BH}$–$M_\mathrm{\ast}$ relation (as otherwise the same large offset would be seen in the $M_\mathrm{BH}$–$M_\mathrm{dyn}$ plane), on the other hand, and may indicate that the $M_\mathrm{BH}$–$M_\mathrm{dyn}$ relation is more fundamental \citep{Maiolino:2024ac}.

In this work, we take a semi-empirical approach to understand the origin of overmassive BHs in the early Universe. We analyze galaxies from the \thzoom simulations \citep{Kannan:2025aa}, which are high-resolution simulations with a state-of-the-art resolved interstellar medium (ISM) model. Importantly, the \thzoom simulations do not include a live BH model, so we calculate the expected BH masses for each galaxy assuming that the local $M_\mathrm{BH}$–$M_\mathrm{dyn}$ relation is fundamental. Using this assumption, we aim to investigate, even without any input BH physics, whether overmassive BHs naturally arise from the evolution of the $M_\mathrm{\ast}$–$M_\mathrm{dyn}$ relation. We show that overmassive BHs result from standard galaxy formation physics and a fundamental $M_\mathrm{BH}$–$M_\mathrm{dyn}$ relation, which has important implications for the early co-evolution of galaxies and their BHs.

This work is organized as follows. In Section~\ref{sec:Methods} we outline the \thzoom simulations and method to calculate BH masses. In Section~\ref{sec:Overmassive black holes}, we show the BH mass distribution and its relation to baryonic and dark matter fractions. In Section~\ref{sec:Discussion and Conclusions} we present our conclusions.

\begin{figure*}
\centering
	\includegraphics[width=\textwidth]{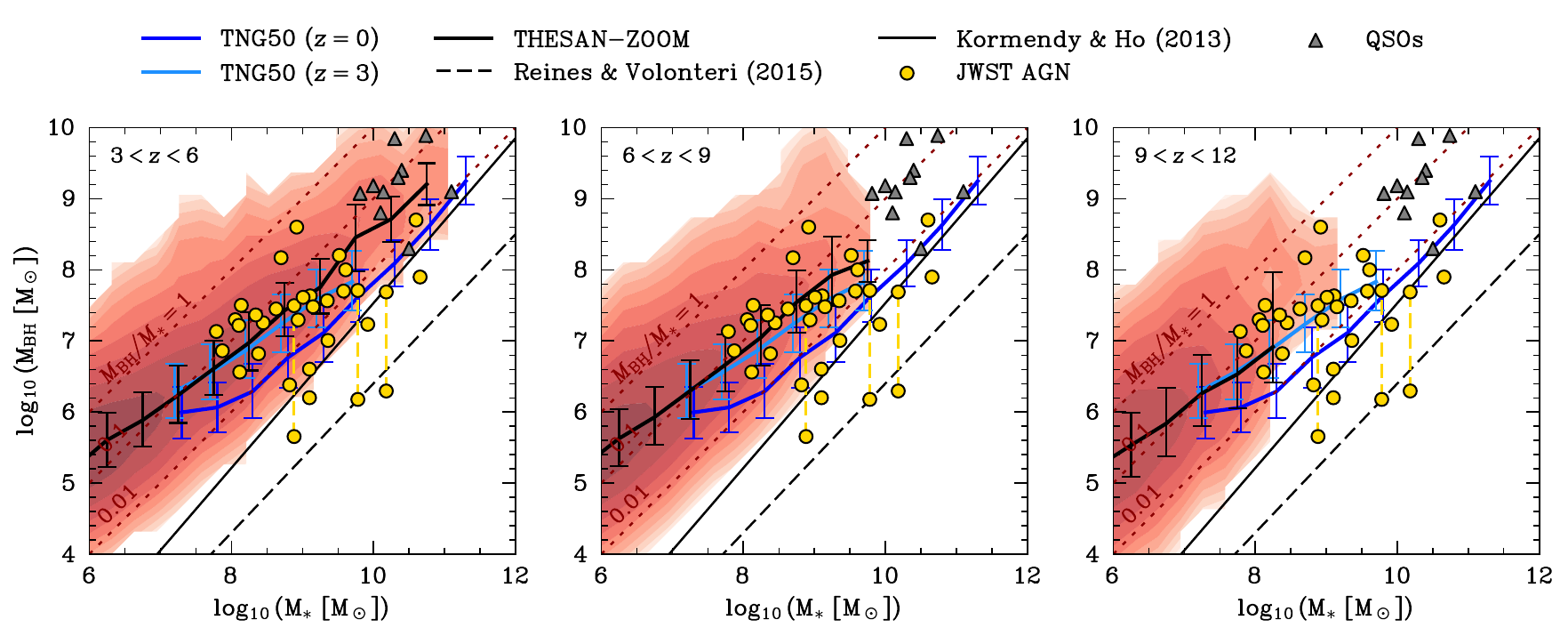}
    \caption{The masses of BHs compared to their host galaxies, where $M_\mathrm{BH}$ is calculated from $M_\mathrm{dyn}$ assuming a universal $M_\mathrm{BH}$–$M_\mathrm{dyn}$ relation (see text for details). Red contours show the log-scaled distribution of \thzoom galaxies. The solid blue line shows the median trend, whereby the $M_\mathrm{BH}/M_\mathrm{\ast}$ ratio evolves from $\sim$0.1 at $M_\mathrm{\ast}=10^6~\mathrm{M_\odot}$ to $\sim$0.01 at $M_\mathrm{\ast}=10^{10.5}~\mathrm{M_\odot}$. Blue errorbars show the 16th--84th percentile scatter. Golden circles show $M_\mathrm{BH}$ for \textit{JWST}-observed AGN at $z\approx4-11$ \citep{Ubler:2023aa,Ubler:2024ab,Harikane:2023aa,Kokorev:2023aa,Carnall:2023aa,Maiolino:2024aa,Maiolino:2024ac,Furtak:2024aa,Juodzbalis:2024ab,Natarajan:2024aa}, with dashed golden lines connecting candidate merging BHs. Grey triangles show QSOs at $z\approx5-7$ \citep{Ding:2022aa,Stone:2024aa,Yue:2024aa}. Dashed red lines follow constant $M_\mathrm{BH}/M_\mathrm{\ast}$ of 1, 0.1, and 0.01. We show best fit relations of $M_\mathrm{BH}$–$M_\mathrm{bulge}$ for local early type galaxies \citep[solid black;][]{Kormendy:2013aa} and $M_\mathrm{BH}$–$M_\mathrm{\ast}$ for local disk galaxies \citep[dashed black;][]{Reines:2015aa}. Even without including any BH physics, our approach well reproduces \textit{JWST}-observed BH and host galaxy masses, including overmassive BHs at low $M_\mathrm{\ast}$. We plot results following the same method for low-mass and bulge-dominated galaxies in TNG50, which show a similar trend at $z=3$ and evolve to the local elliptical relation at $z=0$.
    }
    \label{fig:stellar_mass_vs_bh_mass_contour}
\end{figure*}

\section{Methods}
\label{sec:Methods}

\subsection{The \thzoom project}
\label{sec:Simulations}

This work employs the high-resolution zoom-in \thzoom simulations. A full description of the \thzoom simulations is reported in \citet{Kannan:2025aa}, however we outline the key features here. The \thzoom simulations were carried out with the {\sc arepo-rt} code \citep{Kannan:2019aa}, a radiation hydrodynamics extension of the moving-mesh code {\sc arepo} \citep{Springel:2010aa}. These simulations employ a state-of-the-art model designed to resolve the multi-phase ISM \citep{Marinacci:2019aa,Kannan:2020aa,Zier:2024aa}, allowing us to study processes within galaxies, which are generally not well treated within large-scale cosmological simulations employing an effective equation-of-state approach to the ISM \citep{Pillepich:2018aa}. The \thzoom simulations employ a ``zoom-in'' technique to re-simulate regions of the \thesan parent volume \citep{Kannan:2022aa, Smith:2022ab, Garaldi:2022aa,Garaldi:2024aa}. When inflowing, the time-varying radiation field from the parent volume is injected at the edge of the zoom-in region, allowing radiative feedback from neighboring galaxies to impact the \thzoom galaxies. \thzoom simulations have been used to study the high-redshift star-forming main sequence and burstiness \citep{McClymont:2025aa}, galaxy-scale star-formation efficiencies \citep[SFEs;][]{Shen:2025aa}, the impact of reionisation on galaxies \citep{Zier:2025aa}, the formation of Population III stars \citep{Zier:2025ab}, galaxy size evolution \citep{McClymont:2025ab}, and cloud-scale SFEs \citep{Wang:2025aa}.

Halo catalogs were generated with the friends-of-friends (FOF) algorithm \citep{Davis:1985aa}. Self-gravitating subhalos were identified within the FOF groups with the SUBFIND-HBT algorithm \citep{Springel:2001aa,Springel:2021aa}. In this work, we only consider central galaxies, which we define as the most massive galaxy within a FOF group. 

This high-resolution, multi-phase ISM model is advantageous for this study because it likely more accurately models the structure of high-redshift galaxies \citep[e.g.,][]{Roper:2023aa,McClymont:2025ab}. Additionally, the lack of a live black hole model allows us to test our hypothesis under the assumption that the BHs have a negligible impact on the host galaxy.

\subsection{Inferring black hole masses in \thzoom}
\label{sec:Black hole masses}

The \thzoom simulations do not include a live BH model. We therefore calculate $M_\mathrm{BH}$ by hand using galaxy properties and observed scaling relations. Selecting $M_\mathrm{BH}$ based on the $M_\mathrm{BH}$–$M_\mathrm{\ast}$ relation is inappropriate, given that we are investigating its evolution. Instead, we assume that the relationship between $M_\mathrm{BH}$ and $M_\mathrm{dyn}$ is fundamental. The $M_\mathrm{BH}$–$M_\mathrm{dyn}$ relation is robust even for \textit{JWST}-observed AGN which are overmassive relative to their $M_\mathrm{\ast}$ \citep{Maiolino:2024ac}. The $M_\mathrm{BH}$–$M_\mathrm{dyn}$ relation (and the related $M_\mathrm{BH}$-velocity dispersion relation) may arise due to the accretion rate being modulated by the bulge gravitational potential and BH feedback, where the BH grows until it is massive enough to launch winds which clear surrounding gas \citep{King:2015aa,Marsden:2020aa}.

We measure $M_\mathrm{dyn}$ by summing the stellar, gas, and dark mass within the stellar half-mass radius and multiplying by two, approximating the observationally traced $M_\mathrm{dyn}$. We calculate $M_\mathrm{BH}$ from the $M_\mathrm{BH}$–$M_\mathrm{bulge}$ relation of \citet{Kormendy:2013aa}
\begin{equation}
\label{eq:kh13}
\frac{M_\mathrm{BH}}{10^9\,\mathrm{M_\odot}}=0.49\left( \frac{M_\mathrm{bulge}}{10^{11}\,\mathrm{M_\odot}} \right) ^{1.16} \,,
\end{equation}
where, as in \citet{Maiolino:2024ac}, we are assuming that $M_\mathrm{bulge}\approx M_\mathrm{dyn}$ in the elliptical galaxies on which their relation is based, as they have very little gas content. We neglect the dark matter contribution to $M_\mathrm{dyn}$, which may contribute $\sim$10--40\% in early type galaxies \citep{Cappellari:2006aa,Cappellari:2013aa,Gerhard:2013aa}. A relation calibrated including this contribution could decrease the calculated $M_\mathrm{BH}$ by $\sim$0.05--0.2\,dex. There is an intrinsic scatter of 0.29\,dex in this relation \citep{Kormendy:2013aa}, and thus we randomly draw the BH masses 1000 times from a $\sigma=0.29\,\mathrm{dex}$ Gaussian distribution around the relation. We verified that there are no systemic changes in trend and only a moderate increase in scatter by sampling from the intrinsic scatter. We use the same $M_\ast$ for galaxies as we use in the dynamical mass calculation (i.e., twice the stellar mass within the stellar half-mass radius, or, equivalently, the total stellar mass). We note that if we instead use the masses within twice the stellar half-mass radius for $M_\ast$ and for calculating $M_\mathrm{dyn}$, the resulting $M_\mathrm{BH}$–$M_\mathrm{\ast}$ is increased by $\sim$0.2--0.3\,dex.

\subsection{Observations and SED fitting}
\label{sec:Observations and SED fitting}

To understand the plausibility of the gas fractions we find in simulated galaxies, we infer gas fractions for a sample of observed high-redshift galaxies. To infer these gas fractions we use fits first presented in \citet{Simmonds:2024ab}, which are derived with the spectral energy distribution (SED) fitting code \texttt{Prospector} \citep{Johnson:2019aa,Johnson:2021aa}, on the full \textit{JWST} Advanced Deep Extragalactic Survey \citep[JADES;][]{Eisenstein:2023aa,Eisenstein:2023ab,Bunker:2024aa,DEugenio:2025aa} photometry set in the Great Observatories Origins Deep Survey North and South \citep{Giavalisco:2004aa}. We calculate gas masses using the \citet{Tacconi:2018aa} relation, which gives $M_\mathrm{gas}/M_\ast$ as a function of the specific star-formation rate (sSFR), $M_\ast$, and $z$. We use their $\beta=2$ fit with a \citet{Speagle:2014aa} star-forming main sequence (SFMS), and we calculate the offset from the SFMS using the same SFMS. The \citet{Tacconi:2018aa} relation is calibrated on galaxies with $\log(M_\ast\,\mathrm{[M_\odot]})=9.0-11.8$ at $0<z<4$, and therefore measurements of the gas mass outside this range are extrapolations.

\section{Overmassive black holes}
\label{sec:Overmassive black holes}

\subsection{The black hole--stellar mass relation}
\label{sec:The black hole-stellar mass relation}

\begin{figure*} 
\centering
	\includegraphics[width=\textwidth]{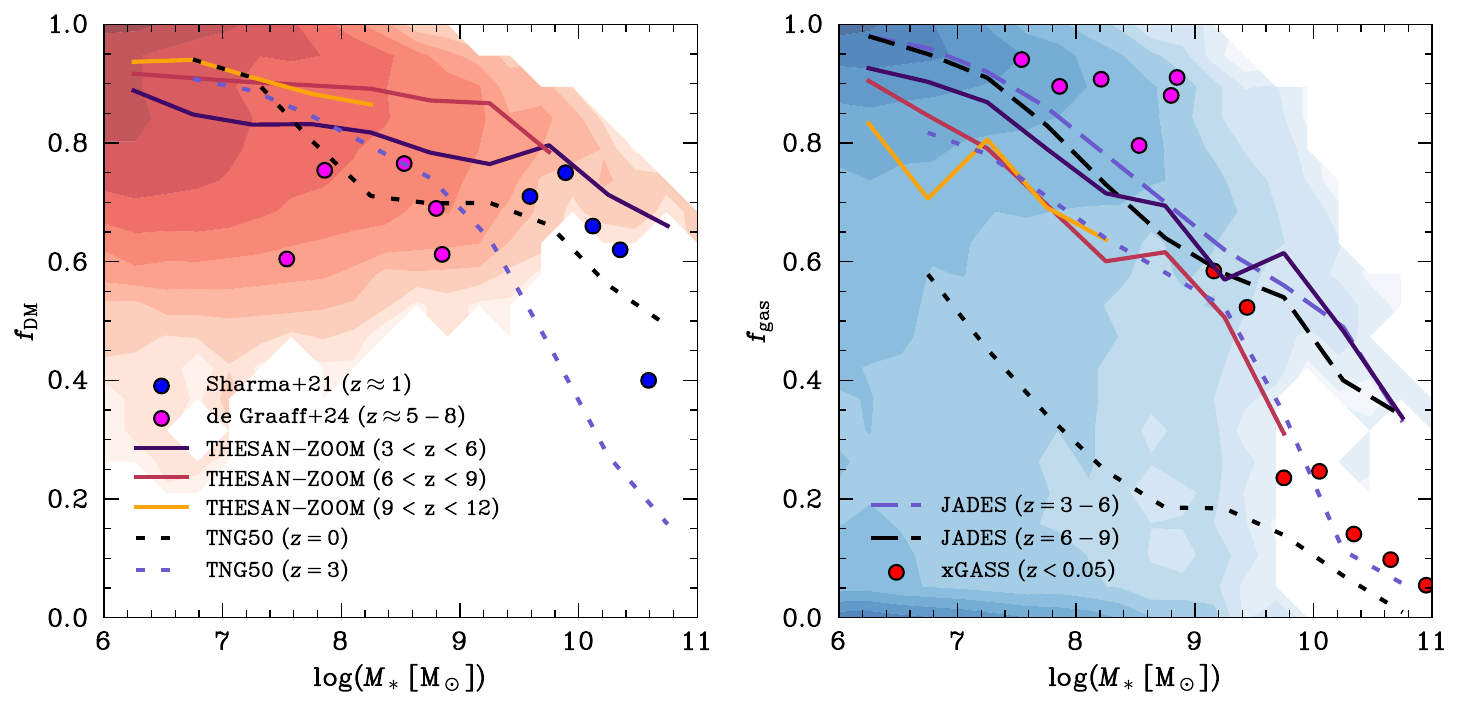}
    \caption{Baryonic and dark matter mass content of galaxies as a function of stellar mass. Contours show the log-scaled distribution of \thzoom galaxies and the lines show median trends. \textit{Left:} $f_\mathrm{DM}$ decreases with $M_\mathrm{\ast}$. \thzoom galaxies have higher $f_\mathrm{DM}$ than the high-redshift observational results of \citet{de-Graaff:2024aa}, this may be due to their overestimated gas masses. Our $f_\mathrm{DM}$ values are consistent with the $z\approx1$ sample of \citet{Sharma:2021aa}, except for the highest mass bin at $M_\mathrm{\ast}\gtrsim10^{10.5}$. There is a weak trend of increasing $f_\mathrm{DM}$ with redshift. TNG50 at $z=0$ shows somewhat lower $f_\mathrm{DM}$ at higher $M_\mathrm{\ast}$, whereas $f_\mathrm{DM}$ is significantly lower for higher masses at $z=3$. \textit{Right:} $f_\mathrm{gas}$ decreases with $M_\mathrm{\ast}$, showing agreement with JADES galaxies. We also show $f_\mathrm{gas}$ from \citet{de-Graaff:2024aa}, which are higher than both JADES and the low-redshift xGASS survey \citep{Catinella:2018aa}. TNG50 shows lower $f_\mathrm{gas}$ for all $M_\mathrm{\ast}$ at $z=0$ and lower $f_\mathrm{gas}$ for higher $M_\mathrm{\ast}$ ($M_\mathrm{\ast}\gtrsim10^{9.5}$) at $z=3$. The median lines show that \thzoom $f_\mathrm{gas}$ values tend to be \textit{higher} at lower redshifts, contrary to expectations. We attribute this to a population of galaxies with $f_\mathrm{gas}\approx0$ at higher redshift, which is caused by galaxies ejecting their ISMs during the burst-quench cycle. The median lines, excluding this population, show no redshift evolution and lie on the plotted $3<z<6$ line, in agreement with the lack of evolution in JADES. The combination of decreasing $f_\mathrm{DM}$ and $f_\mathrm{gas}$ with $M_\mathrm{\ast}$ leads to increasing $M_\mathrm{\ast}/M_\mathrm{dyn}$ with $M_\mathrm{\ast}$.}
    \label{fig:joint_fractions}
\end{figure*}

In Fig.~\ref{fig:stellar_mass_vs_bh_mass_contour} we show the distribution of BH masses as a function of stellar mass in three redshift bins. There is no clear redshift evolution at a fixed stellar mass down to $z=3$, although the scatter does seem to increase with redshift. Our BH masses cover the distribution of those observed with \textit{JWST} in low stellar mass ($M_\mathrm{\ast}\lesssim10^9\,\mathrm{M_\odot}$) galaxies at $z>3$. Importantly, low stellar mass galaxies show significantly more overmassive BHs relative to the local $M_\mathrm{BH}$–$M_\mathrm{\ast}$ relation compared to their higher-mass counterparts, with the median ratio evolving from $M_\mathrm{BH}/M_\mathrm{\ast}\approx0.1$ at $M_\mathrm{\ast}=10^6\,\mathrm{M_\odot}$ to $M_\mathrm{BH}/M_\mathrm{\ast}\approx0.01$ at $M_\mathrm{\ast}=10^{10.5}\,\mathrm{M_\odot}$. This trend is in general agreement with observations, where the locally calibrated $M_\mathrm{BH}/M_\mathrm{\ast}$ relation is seen to fail more dramatically for lower stellar masses \citep{Maiolino:2024ac}, although it should be noted that there are few constraints locally below $M_\mathrm{\ast}\lesssim10^{9.5}\,\mathrm{M_\odot}$. Still, at higher stellar masses ($M_\mathrm{\ast}\gtrsim10^{10}\,\mathrm{M_\odot}$), our calculated BH masses fall in the region occupied by quasars. 

The \citet{Kormendy:2013aa} relation sets the minimum BH mass at a given stellar mass for our semi-empirical approach, as this is the BH mass we would calculate even if the only contribution to the dynamical mass is from stars (i.e., the galaxy is devoid of gas and dark matter). That the \citet{Kormendy:2013aa} relation sets a rough lower limit means that our agreement at higher stellar masses is difficult to judge. This is further complicated by the choice of radius within which to compute $M_\mathrm{dyn}$. For local galaxies, scaling relations between BHs and their host galaxies are seen to be strongest when only the spheroidal or bulge component is used \citep{Kormendy:2013aa}. We have implicitly assumed that the stellar-half mass radius exclusively covers a spheroid-like component, and while this is likely valid for most of the \thzoom galaxies, it may break down for the most massive galaxies at $z=3$ if an extended disk begins to form, leading us to overestimate the BH masses for the most massive galaxies. 

To provide a direct comparison, we apply our semi-empirical method to the TNG50 simulations \citep{Nelson:2019ab,Pillepich:2019aa}. We explicitly ignore the BHs in TNG50 and instead follow the same procedure as for \thzoom to calculate $M_\mathrm{BH}$. We only include central galaxies with $M_\ast>10^{7}\,\mathrm{M_\odot}$. There is a publicly available catalog of morphological decompositions for TNG50 for galaxies with $M_\ast>10^{9}\,\mathrm{M_\odot}$ \citep{Zana:2022aa}. While this catalog does not give the information needed to calculate the dynamical masses of the bulge, it does allow us to restrict our sample to bulge-dominated galaxies, where the thick and thin disks comprise less than 25\% of the stellar mass. At $z=3$, TNG50 shows a comparable relation to \thzoom, whereas by $z=0$ the inferred BH masses have decreased rapidly. At higher masses at $z=0$, bulge-dominated galaxies in TNG50 scatter around the local elliptical relation, as expected.

\subsection{Baryonic and dark mass content}
\label{sec:Baryonic and dark mass content}

\begin{table*}
    \centering
    \renewcommand{\arraystretch}{1.15}
    \begin{tabular}{c cc cc cc c c}
        \hline
        & \multicolumn{2}{|c|}{\thzoom} & \multicolumn{2}{|c|}{\thzoom} & \multicolumn{2}{|c|}{\thzoom} & \multicolumn{1}{|c|}{JADES} & \multicolumn{1}{|c|}{JADES}\\
        & \multicolumn{2}{|c|}{$3<z<6$} & \multicolumn{2}{|c|}{$6<z<9$} & \multicolumn{2}{|c|}{$9<z<12$} & \multicolumn{1}{|c|}{$3<z<6$} & \multicolumn{1}{|c|}{$6<z<9$}\\
        \hline
        $\log(M_\mathrm{\ast}\,[\mathrm{M_\odot}])$& $f_\mathrm{DM}$ & $f_\mathrm{gas}$ & 
        $f_\mathrm{DM}$ & $f_\mathrm{gas}$ & 
        $f_\mathrm{DM}$ & $f_\mathrm{gas}$ & 
        $f_\mathrm{gas}$ & 
        $f_\mathrm{gas}$ \\
        6.25 & $0.89^{+0.09}_{-0.13}$ & $0.93^{+0.05}_{-0.89}$ & $0.92^{+0.07}_{-0.13}$ & $0.9^{+0.07}_{-0.9}$ & $0.94^{+0.05}_{-0.17}$ & $0.83^{+0.14}_{-0.83}$ & $0.98^{+0.01}_{-0.01}$ & $0.98^{+0.01}_{-0.01}$ \\
        6.75 & $0.85^{+0.12}_{-0.13}$ & $0.9^{+0.06}_{-0.79}$ & $0.91^{+0.07}_{-0.13}$ & $0.85^{+0.11}_{-0.85}$ & $0.94^{+0.04}_{-0.16}$ & $0.71^{+0.25}_{-0.71}$ & $0.96^{+0.01}_{-0.01}$ & $0.95^{+0.01}_{-0.01}$ \\
        7.25 & $0.83^{+0.13}_{-0.14}$ & $0.87^{+0.08}_{-0.83}$ & $0.9^{+0.07}_{-0.13}$ & $0.79^{+0.15}_{-0.79}$ & $0.91^{+0.06}_{-0.14}$ & $0.81^{+0.13}_{-0.81}$ & $0.92^{+0.01}_{-0.01}$ & $0.91^{+0.01}_{-0.01}$ \\
        7.75 & $0.83^{+0.12}_{-0.14}$ & $0.79^{+0.13}_{-0.75}$ & $0.9^{+0.06}_{-0.13}$ & $0.7^{+0.21}_{-0.7}$ & $0.88^{+0.06}_{-0.08}$ & $0.69^{+0.22}_{-0.69}$ & $0.86^{+0.01}_{-0.01}$ & $0.83^{+0.01}_{-0.01}$ \\
        8.25 & $0.82^{+0.1}_{-0.13}$ & $0.72^{+0.16}_{-0.67}$ & $0.89^{+0.05}_{-0.11}$ & $0.6^{+0.26}_{-0.6}$ & $0.86^{+0.07}_{-0.04}$ & $0.64^{+0.29}_{-0.64}$ & $0.78^{+0.01}_{-0.01}$ & $0.73^{+0.01}_{-0.01}$ \\
        8.75 & $0.78^{+0.1}_{-0.1}$ & $0.69^{+0.15}_{-0.45}$ & $0.87^{+0.06}_{-0.1}$ & $0.62^{+0.23}_{-0.59}$ &  &  & $0.7^{+0.01}_{-0.01}$ & $0.64^{+0.01}_{-0.01}$ \\
        9.25 & $0.76^{+0.09}_{-0.09}$ & $0.57^{+0.17}_{-0.41}$ & $0.87^{+0.03}_{-0.07}$ & $0.51^{+0.23}_{-0.45}$ &  &  & $0.62^{+0.01}_{-0.01}$ & $0.58^{+0.02}_{-0.02}$ \\
        9.75 & $0.8^{+0.05}_{-0.08}$ & $0.61^{+0.14}_{-0.22}$ & $0.79^{+0.03}_{-0.04}$ & $0.31^{+0.17}_{-0.3}$ &  &  & $0.56^{+0.01}_{-0.01}$ & $0.54^{+0.04}_{-0.03}$ \\
        10.25 & $0.71^{+0.05}_{-0.04}$ & $0.48^{+0.09}_{-0.21}$ &  &  &  &  & $0.49^{+0.01}_{-0.01}$ & $0.4^{+0.05}_{-0.07}$ \\
        10.75 & $0.66^{+0.05}_{-0.06}$ & $0.34^{+0.08}_{-0.17}$ &  &  &  &  & $0.33^{+0.03}_{-0.03}$ & $0.34^{+0.01}_{-0.09}$ \\
        
        \hline
    \end{tabular}
    \caption{Gas and dark matter fractions for \thzoom galaxies in 0.5\,dex stellar mass bins. The errors on the simulation gas fractions represent the $16^\mathrm{th}-84^\mathrm{th}$ percentile scatter. Bins are only included where we have at least 10 galaxies. We also show the median gas fractions for JADES galaxies, derived from our \textsc{Prospector} SED fitting and the \citet{Tacconi:2018aa} gas fraction scaling relation. The errors on JADES gas fractions are the bootstrap uncertainties, resampling from measurement errors in SFR, $M_\ast$, and $z$ with replacement 1000 times. We enforce a minimum error of 1\%.}
    \label{tab:frac}
\end{table*}

Given BH masses are calculated directly from the dynamical masses of our galaxies, it is simple to identify the cause of the extreme $M_\mathrm{BH}/M_\mathrm{\ast}$ ratios at low stellar masses: the dynamical masses of low stellar mass galaxies must be dominated by non-stellar matter. In the left panel of Fig.~\ref{fig:joint_fractions} we show the dark matter fraction, $f_\mathrm{DM}$, as a function of stellar mass in three redshift bins. The dark matter fraction is calculated within the stellar half-mass radius as $f_\mathrm{DM}=M_\mathrm{DM}/(M_\mathrm{DM}+M_\mathrm{gas}+M_\mathrm{\ast})$. The evolution of $f_\mathrm{DM}$ with redshift appears weak at a fixed stellar mass, which agrees with the weak, if any, redshift evolution of BH mass shown in Fig.~\ref{fig:stellar_mass_vs_bh_mass_contour}. The dark matter fraction decreases with increasing stellar mass, which is in general agreement with the observed trends, although our dark matter fractions are notably $\sim$0.2 higher than those from \citet{de-Graaff:2024aa,de-Graaff:2024ab}. This does not necessarily indicate an issue with the \thzoom dark matter fractions as the sample of \citet{de-Graaff:2024aa,de-Graaff:2024ab} comprises only six galaxies (including one galaxy with nonphysical $f_\mathrm{DM}<0$), and given that the gas mass in these galaxies may be overestimated (see the high gas fraction in the right panel of Fig.~\ref{fig:joint_fractions}), which would systemically bias the dark matter fractions low.

In the right panel of Fig.~\ref{fig:joint_fractions} we show the median gas fraction, $f_\mathrm{gas}$, of \thzoom galaxies. The values are provided in Tab.~\ref{tab:frac}. The gas fraction is also calculated within the stellar half-mass radius as $f_\mathrm{gas}=M_\mathrm{gas}/(M_\mathrm{gas}+M_\mathrm{\ast})$. Similarly to the dark matter fraction, the gas fraction decreases with stellar mass, however, it shows a noticeably sharper decrease. The median trend at $3<z<6$ is in good agreement with the observationally derived gas fractions, however, the trends indicate that the gas fractions decrease with redshift, contrary to our expectation and to the observational results, which show no evidence of redshift evolution from $z=3$ to $z=9$. This apparent redshift evolution is exclusively due to a population of galaxies with $f_\mathrm{gas}\approx0$. If we calculate the median trends while excluding galaxies with $f_\mathrm{gas}<0.2$, then the redshift evolution disappears and the median trend for each redshift bin lies on the $3<z<6$ line shown in Fig.~\ref{fig:joint_fractions}. We attribute galaxies with $f_\mathrm{gas}\approx0$ to those that have ejected their ISM during the burst-quench cycle. Galaxies in this state are more prevalent at higher redshifts due to increasingly bursty star formation \citep{McClymont:2025aa}. Low mass galaxies with extremely low gas fractions are generally not observable due to their low SFR, which explains why such galaxies are not present in our analysis of JADES galaxies, although ``mini-quenched'' galaxies, which show evidence of star formation but little nebular emission, could plausibly have such low gas fractions \citep{Looser:2024aa}.

Interestingly, in both JADES and \thzoom galaxies (excluding galaxies with extremely low gas fractions), $f_\mathrm{gas}$ shows no clear evidence of redshift evolution at $z>3$. In Fig.~\ref{fig:joint_fractions}, we also show $f_\mathrm{gas}$ from the xGASS survey of local galaxies. For galaxies with $M_\ast\lesssim10^{9.5}\,\mathrm{M_\odot}$, the local gas fractions agree with those we measure at $z>3$ in JADES and in the \thzoom simulation. However, at $M_\ast\gtrsim10^{9.5}\,\mathrm{M_\odot}$ we find significant disagreement, with $f_\mathrm{gas}\approx0.6$ in JADES and \thzoom at $M_\ast=10^{10}\,\mathrm{M_\odot}$, compared to $f_\mathrm{gas}\approx0.25$ in xGASS. It appears that higher mass galaxies show significant evolution at $z<3$, whereas low mass galaxies do not. This is plausibly due to the increased fraction of quenched galaxies at $z\approx0$, which are predominantly more massive and have low gas fractions. To gain insight into the evolution to $z=0$, we show $f_\mathrm{gas}$ and $f_\mathrm{DM}$ from TNG50. We followed the same procedure as for \thzoom, and we do not enforce a bulge-dominated selection for this comparison. TNG50 shows consistently lower $f_\mathrm{gas}$ across the stellar mass range, and lower $f_\mathrm{DM}$ at higher stellar masses, indicating evolution of the $M_\ast-M_\mathrm{dyn}$ relation. We note, however, that $f_\mathrm{DM}$ is aperture-sensitive, and even tends to increase at extremely high stellar masses ($M_\ast\gtrsim10^{11}\,\mathrm{M_\odot}$) in TNG \citep{Lovell:2018aa}. To calculate $M_\mathrm{BH}$ for TNG50 galaxies, an even smaller aperture would likely be appropriate to reflect the bulge radius, which would further reduce $f_\mathrm{DM}$, particularly at higher stellar masses. Indeed, \citet{de-Graaff:2024ab} find high-mass $z=0$ galaxies are significantly more baryon-dominated using a fixed 1\,kpc aperture in TNG50. Interestingly, at $z=3$, TNG50 shows similar $f_\mathrm{DM}$ and $f_\mathrm{gas}$ as \thzoom, except at $M_\ast\gtrsim10^{9.5}\,\mathrm{M_\odot}$, where both are lower.

We now consider whether both high $f_\mathrm{DM}$ and high $f_\mathrm{gas}$ are crucial for producing overmassive BHs, or whether either is sufficient to produce overmassive BHs alone. If we recalculate our $M_\mathrm{BH}$ vs $M_\mathrm{\ast}$ relation (shown in Fig.~\ref{fig:stellar_mass_vs_bh_mass_contour}) but now assume that gas is instantly converted to stars, we find that the median trends fall by $\sim$1\,dex as galaxies are shifted to the right of the plot. However, given that $\sim$90\% of the dynamical mass contribution is from dark matter, if the systems were instead baryon dominated (i.e., the dark matter was removed), then the BH masses would also fall by $\sim$1\,dex. Therefore, we conclude that both high $f_\mathrm{DM}$ and $f_\mathrm{gas}$ are required to generate a distribution of dynamical and stellar masses that reproduces observed BH masses.

\section{Discussion and Conclusions}
\label{sec:Discussion and Conclusions}

We have shown that we can reproduce the observed distribution of BH masses at $z>3$ as a function of stellar mass by assuming that the locally measured $M_\mathrm{BH}$–$M_\mathrm{dyn}$ relation is fundamental. With this assumption, the overmassive properties of BHs on the $M_\mathrm{BH}$–$M_\mathrm{\ast}$ diagram arise because low-mass galaxies are more gas-rich and dark matter-dominated compared to more massive galaxies, and so they show low $M_\mathrm{\ast}/M_\mathrm{dyn}$. An interesting implication of our results is that we expect ratios of $M_\mathrm{BH}/M_\mathrm{\ast}\approx0.05$ to be typical for high-redshift galaxies. This implies that BHs grew rapidly in the earliest stages of galaxy formation, and potentially had an important impact on their nascent host galaxies. We note that our semi-empirical model does not take into account the seeding of BHs. However, a fundamental $M_\mathrm{BH}$–$M_\mathrm{dyn}$ relation implies that, regardless of seed mass, the BH would quickly accrete material to lie on this relation (within some intrinsic scatter), although we highlight that this early growth is an assumption of our approach and that we cannot model it directly.

If the $M_\mathrm{BH}$–$M_\mathrm{dyn}$ relation is indeed fundamental, it is curious that such a relation does not track the stellar mass-halo mass relation. This implies that in the earliest stages of galaxy formation, gas can collapse sufficiently to accrete onto the central BH, but not to form a commensurate number of stars to produce a comparably high stellar to BH mass ratio. The divergence of stellar and BH mass indicates fundamental differences in the regulation of star formation and BH accretion, which become apparent at low masses and early times. While our assumption of a fundamental $M_\mathrm{BH}$–$M_\mathrm{dyn}$ relation appears consistent with observations thus far, that observed BH masses may be revised in the future. However, we note that if BH masses are revised to match the local $M_\mathrm{BH}$–$M_\mathrm{\ast}$ relation, this would require that the $M_\mathrm{BH}$–$M_\mathrm{dyn}$ breaks down, which itself would have important implications for the early growth of BHs.

While the \thzoom simulations terminate at $z=3$, it is interesting to consider the implications for BHs in local dwarf galaxies. Assuming a fundamental $M_\mathrm{BH}$–$M_\mathrm{dyn}$ relation, we would expect overmassive BHs in local gas-rich dwarfs, similar to their higher redshift counterparts. Confirmation of AGN in local dwarf galaxies is challenging due to the faintness of broad lines and the concern for contamination by supernovae and stellar winds \citep{Izotov:2007aa}. Regardless, among the candidates sources reported in the literature, there are indeed indications of overmassiveness at $M_*<10^{9.5}~M_\odot$. For example, among the four local dwarf AGN candidates reported by \citet{Izotov:2008aa}, one has $M_\mathrm{BH}/M_\mathrm{\ast}\approx 0.026$ at $\log (M_\ast\,\mathrm{[M_\odot]})= 8.4\pm 0.4$, despite the large uncertainties in its stellar mass \citep{Burke:2021aa}. The other three candidates have $M_\ast=10^{9-10}~\mathrm{M_\odot}$ and are more consistent with the local relation. In addition, \citet{Hatano:2024aa} reported an overmassive dwarf AGN candidate, J1205, with $M_\mathrm{BH}/M_\mathrm{\ast}\approx 0.022$ at $\log (M_\ast\,[\mathrm{M_\odot}]) \approx 6.84$, and the potential overmassiveness of the dwarf AGN candidate, SBS 0335-052E, at $\log (M_\ast\,\mathrm{[M_\odot]}) \approx 7.7$, and interestingly this AGN shows Balmer line absorption, which is similar to many \textit{JWST}-observed AGN \citep{Trefoloni:2025aa}. More recently, \citet{Lin:2025aa} present a sample of 19 broad-line AGN selected from local green pea galaxies at $z\approx 0.1-0.3$, which show ubiquitous overmassiveness compared to the local relation with $M_\mathrm{BH}/M_\mathrm{\ast}$ up to 0.04.

On the other hand, many local dwarfs are gas-poor, which presents interesting prospects for the prediction of the masses of their BHs depending on their star-formation histories, assuming that the $M_\mathrm{BH}$–$M_\mathrm{dyn}$ relation was fundamental as these galaxies grew. If most of the gas from these galaxies was ejected during the last star-formation event (i.e., the formation of the last stars before the galaxy became quiescent), rather than converted into stars, then we may expect their BHs to be overmassive even relative to their dynamical mass, as in this case it would be impossible for them to obey the $M_\mathrm{BH}$–$M_\mathrm{dyn}$ relation at all times. However, if most of their gas was converted into stars and accreted onto their BH, we would expect them to lie on the $M_\mathrm{BH}$–$M_\mathrm{dyn}$ relation. This scenario, whereby star formation in dwarfs ceased primarily via starvation due to reionization, is perhaps more likely \citep[e.g.][]{Katz:2020aa,Rey:2025aa}. In either case, comparison of BH masses between gas-rich and gas-poor dwarfs would be illuminating. 

Finally, we must emphasize the limitations of our approach by noting that the lack of a live BH model prevents us from directly exploring the co-evolution of galaxies and their BHs. While it is remarkable that observations can be reproduced without live BH physics, such overmassive black holes could strongly impact the stellar and gas content of these galaxies via feedback, which we cannot explore with our approach. Our results essentially rely on the assumption that feedback from these black holes is inefficient relative to stellar feedback. A more complete and self-consistent understanding requires approaches that can more directly explore the co-regulation of BH and stellar mass growth at high redshift. The \thzoom model offers a promising avenue to explore this topic given that star formation in these simulations is extremely bursty \citep{McClymont:2025aa} and that these starbursts tend to be centrally concentrated \citep{McClymont:2025ab}, raising the prospect of efficient BH feeding during compact starburst events. We anticipate future studies exploring the co-evolution of galaxies and their BHs in the early Universe through the incorporation of live BH modelling within the \thzoom framework.

\section*{Acknowledgements}

The authors are grateful to the referee for their helpful comments, which improved the manuscript. The authors gratefully acknowledge the Gauss Centre for Supercomputing e.V. (\url{www.gauss-centre.eu}) for funding this project by providing computing time on the GCS Supercomputer SuperMUC-NG at Leibniz Supercomputing Centre (\url{www.lrz.de}), under project pn29we. WM thanks the Science and Technology Facilities Council (STFC) Center for Doctoral Training (CDT) in Data Intensive Science at the University of Cambridge (STFC grant number 2742968) for a PhD studentship. WM and ST acknowledge support from the Royal Society Research Grant G125142.
RM and XJ acknowledge support by the Science and Technology Facilities Council (STFC), by the ERC through Advanced Grant 695671 ``QUENCH'', and by the UKRI Frontier Research grant RISEandFALL. RM also acknowledges funding from a research professorship from the Royal Society. RK acknowledges support of the Natural Sciences and Engineering Research Council of Canada (NSERC) through a Discovery Grant and a Discovery Launch Supplement (funding reference numbers RGPIN-2024-06222 and DGECR-2024-00144) and York University's Global Research Excellence Initiative. XS acknowledges the support from the NASA theory grant JWST-AR-04814.

\section*{Data Availability}

All simulation data, including snapshots, group, and subhalo catalogs and  merger trees will be made publicly available in the near future. Data will be distributed via \url{www.thesan-project.com}. Before the public data release, data underlying this article will be shared on reasonable request to the corresponding author(s).



\bibliographystyle{mnras}
\bibliography{overmassive} 




\appendix


\bsp	
\label{lastpage}
\end{document}